\documentstyle[12pt]{article}

\newcommand{\spone}{0.9}

\newcommand{\singlespace}{\edef\baselinestretch{\spone}\Large\normalsize}

\singlespace
\begin{document}
\singlespace

\begin{center}
{\bf The Hilbert-Schmidt Theorem Formulation of the R-Matrix Theory}  \\
\vspace{24pt}

Yeong E. Kim and Alexander L. Zubarev \\
\vspace{5pt}

Department of Physics, Purdue University \\
\vspace{5pt}

West Lafayette, IN  47907 \\
\end{center}
\vspace{8pt}

\begin{center}
{\bf Abstract}
\end{center}
\vspace{5pt}

Using the Hilbert-Schmidt theorem, we reformulate the R-matrix theory
in terms of a uniformly and absolutely convergent expansion.  Term by term
differentiation is possible with this expansion in the neighborhood
of the surface.  Methods for improving the convergence are discussed
when the R-function series is truncated for practical applications.
\vspace{8pt}

\noindent

{\bf I.  Introduction}
\vspace{5pt}

Since 1947, boundary condition methods (BCM) have played an important role
for many quantum mechanical problems [1-37].  In the BCM formulation,
configuration space is divided into two parts:  internal and external
regions.  In the external region, the interaction is usually known and in
many cases the effective two-body equation is exactly solvable.  A
boundary condition matrix is defined in terms of the independent external
wave functions and their derivates at a boundary.  From this information
(boundary condition matrix) and the known solution in the external region,
the S-matrix and the cross-section can be calculated.   There are two
boundary condition matrices:  R matrix and P matrix.  The R-matrix, which
is the inverse of the logarithmic derivative of the external channel wave
function at the surface, was first introduced by Wigner and Eisenbud in
1947 [1].  A detailed account of the R-matrix theory of nuclear reactions
is given in [3].  The P-matrix is the inverse of the R-matrix.  The
P-matrix formulation of nuclear reactions has not been used extensively
except for the nucleon-nucleon scattering problem [5-10].
\vspace{8pt}

The R-matrix theory is extensively employed for describing energy
dependence of the cross-section for various binary nuclear processes including
both elementary and composite nuclear particles [3, 11-17] and is often
used to extrapolate experimental data for the cross-section which are dominated by the contributions from
a few resonance or bound states.  
\vspace{8pt}

In atomic physics, the exchange interaction, which is the most difficult
part of the problem to calculate accurately, is only in the internal
region, and interactions in the external region reduce to long range
local potentials [18].  The R-matrix code is a very powerful computation
code [19, 20] for calculating electron-atom collisions and photoabsorption
processes.  For example, the inner-shell photoionization cross-sections
calculated by the R-matrix code [21, 22] are in excellent agreement with the
recent experimental measurements [38].  
\vspace{8pt}

The R-matrix method for studying low-energy electron-molecule collisions
was developed in [23-25].  It has been used to describe elastic
scattering, electronic excitation, vibrational excitation [27] and
dissociative attachment [29].
\vspace{8pt}

The R-matrix method is based upon expanding the total wave function $\Psi$
for any energy in the internal region in terms of the complete set of eigenfunction
$X_\lambda$ of that region,
$$ \Psi = \sum_\lambda~A_\lambda X_\lambda  ,
\eqno{(1)}
$$

\noindent
where $X_\lambda$ are defined by the equation
$$ HX_\lambda = E_\lambda X_\lambda ,
\eqno{(2)}
$$

\noindent
and satisfy certain R-matrix boundary conditions on the surface [3].
It is known that there is a formal problem with the wave function
expansion used in the conventional approach for the R-matrix theory.
Either the expansion given by Eq. (1) is not uniformly convergent
in the neighborhood
of the surface, or term-by-term differentiation of the expansion is not
admissible [3,18].  To avoid these difficulties variational formalisms
[30-36] were proposed
with basis functions which do not satisfy the R-matrix
boundary conditions.
\vspace{8pt}

The R-matrix theory is rigorous and therefore, there
remains, from the formal point of view, a problem of how to formulate
the R-matrix method in terms of a uniformly and absolutely convergent
expansion.
\vspace{8pt}

In this paper we present solutions of this formal problem and discuss
various approximations of the R-function.  In Section II, we describe
in some detail the conventional formulation of R-function in terms of the
expansion given by Eq. (1) and (2).  In Section III, we reformulate
the R-matrix theory based on the Hilbert-Schmidt theorem to obtain the
R-function in terms of a series which is uniformly and absolutely 
convergent.  In Section IV, we discuss methods of improving the
convergence of the R-function series when it is truncated for practical
applications.  A summary and conclusions are given in Section V. 
\vspace{8pt}

\noindent
{\bf II.  R-function}
\vspace{5pt}

In order to describe the formal procedure employed in the R-matrix
theory, we consider the simplest case of potential scattering for spinless
particles with only the elastic scattering channel being open.
\vspace{8pt}

The radial wave function $u_\ell(r)$ in the interior region $0 \leq r \leq a$
satisfies the Schr\"odinger equation
$$ - \frac{d^2u_\ell}{dr^2} + (\frac{2\mu V(r)}{\hbar^2} + \frac{\ell(\ell +1)}{r^2})
u_\ell(r) = k^2 u_\ell(r),
\eqno{(3)}
$$

\noindent
where $\mu$ is the reduced mass, $V(r)$ is the interaction potential in the internal region $0 \leq r \leq a$,
and $k^2 = 2\mu E/\hbar^2$.
\vspace{8pt}

In the conventional theory [3, 18, 37], $u_\ell(r)$ in the internal region
$(0 \leq r \leq a)$ is expanded in terms of complete set of states
$u_\lambda^\ell (r)$ given within the region $0 \leq r \leq a$.  These states
are the solutions of the equation
$$ - \frac{d^2u_\lambda^\ell}{dr^2} + (\frac{2\mu V(r)}{\hbar^2} +
\frac{\ell(\ell + 1)}{r^2})u_\lambda^\ell(r) = k_\lambda^2 u_\lambda^\ell(r),
\eqno{(4)}
$$

\noindent
satisfying the R-matrix boundary conditions
$$
\everymath={\displaystyle}
\begin{array}{rcl}
 &~& u_\lambda^\ell(0) = 0, \\
 &~& \\
&~& \frac{a}{u_\lambda^\ell(a)} (\frac{du_\lambda^\ell}{dr})_{r=a} ~ = B,
\end{array}
\eqno{(5)}
$$

\noindent
and the orthonormality conditions
$$\int_o^a~u_\lambda^\ell(r)u_{\lambda^\prime}^\ell(r)dr =
\delta_{\lambda\lambda^\prime}.
\eqno{(6)}
$$

In the region $0 \leq r \leq a, ~ u_\ell(r)$ may be expanded in terms of the
eigenfunctions $u_\lambda^\ell(r)$.
$$ u_\ell(r) = \sum_{\lambda = 1}^\infty ~ c_\lambda^\ell 
u_\lambda^\ell(r),~~~~(0 \leq r \leq a),
\eqno{(7)}
$$

\noindent
where
$$c_\lambda^\ell = \int_o^a dru_\ell(r)u_\lambda^\ell(r).
\eqno{(8)}
$$

As we show below, either (i)
the expansion (7) does not converge uniformly, or (ii) term-by-term
differentiation is not admissible [3, 18], or both (i) and (ii) may be applicable.
From Green's theorem [3, 18]
and the boundary conditions (5), we find
$$ c_\lambda^\ell = \frac{1}{a}~\frac{u_\lambda^\ell(a)}{k_\lambda^2- k^2}
[a ~ \frac{du_\ell}{dr} - Bu_\ell]_{r=a} .
\eqno{(9)}
$$

\noindent
Substitution of Eq. (7) into Eq. (6) gives
$$ u_\ell(r)[a \frac{du_\ell}{dr} - Bu_\ell]_{r=a}^{-1}= \frac{1}{a}
\sum_{\lambda = 1}^\infty ~ \frac{u_\lambda^\ell(r) u_\lambda^\ell(a)}{k_\lambda^2 -
k^2} .
\eqno{(10)}
$$

\noindent
If we now define
$$ R^{(B)} = \frac{1}{a} \sum_{\lambda = 1}^\infty
 \frac{(u_\lambda^\ell(a))^2}{k_\lambda^2 - k^2} ,
\eqno{(11)}
$$
 
\noindent
and assume that
$$ [\sum_{\lambda = 1}^\infty ~ \frac{u_\lambda^\ell(r)u_\lambda^\ell(a)}{k_\lambda^2 -
k^2}]_{r=a} = \sum_{\lambda = 1}^\infty~\frac{(u_\lambda^\ell(a))^2}{k_\lambda^2 -
k^2} ,
\eqno{(12)}
$$

\noindent
we find that $R^{(B)}$ relates the amplitude of $u_\ell$ to its derivative at
the boundary by the relation
$$ R^{(B)} = u_\ell(a)[a \frac{du_\ell}{dr} - Bu_\ell]_{r = a}^{-1} .
\eqno{(13)}
$$

\noindent
Once $R^{(B)}$ has been calculated, the K-matrix and cross-section can be easily
determined. 
\vspace{8pt}

From the fact that
$$G_\ell(r,r^\prime) =
- \sum_{\lambda=1}^\infty~\frac{u_\lambda^\ell(r)u_\lambda^\ell(r^\prime)}{k_\lambda^2-k^2} 
\eqno{(14a)}
$$

\noindent
and

$$G_\ell(r,r^\prime) = \Biggr\{ \begin{array}{ll} 
    u_\ell(r)Y_\ell(r^\prime),~~& r \leq r^\prime, \\
    u_\ell(r^\prime)Y_\ell(r),~~& r^\prime \leq r , \\
\end{array} 
\eqno{(14b)}
$$

\noindent
with 

$$ Y_\ell(r) = \frac{y_\ell(a)Bu_\ell(r)}{[a \frac{du_\ell}{dr} - Bu_\ell]_{r=a}} + y_\ell(r),
\eqno{(15)}
$$

\noindent
where $y_\ell(r)$ is the irregular solution of Eq. (3) with boundary conditions
$(\frac{dy_\ell}{dr})_{r=a} = 0$, and $u_\ell^\prime y_\ell - y_\ell^\prime u_\ell = 1$, it can be seen 
that Eqs. (7) and (10) can be obtained from the spectral decomposition, Eq. (14a), of Green's
function $G_\ell(r,r^\prime)$.  This bilinear series, Eq. (14a), converges in
$L_2 (0 < r < a)$. 
\vspace{8pt}

We note that the completeness of the states $u_\lambda^\ell$
does not guarantee validity of Eq. (12).
To demonstrate this statement, let us first consider a special case of
boundary conditions $B = \infty$, or
$$ u_\lambda^\ell(0) = u_\lambda^\ell(a) = 0.
\eqno{(16)}
$$

\noindent
In this case, we can find from Eq. (9) that
$$ c_\lambda^\ell = \frac{u_\ell(a)}{k^2 - k_\lambda^2}~(\frac{du_\lambda^\ell}{dr})_{r=a},
\eqno{(17)}
$$

\noindent
and substitution of Eq. (17) into Eq. (7) gives
$$ \frac{u_\ell(r)}{u_\ell(a)} = \sum_{\lambda = 1}^\infty~\frac{u_\lambda^\ell(r)}{k^2 - k_\lambda^2}(\frac{du_\lambda^\ell}{dr})_{r=a} .
\eqno{(18)}
$$

\noindent
If one tries to obtain this value of $\displaystyle\lim_{r \rightarrow a}~\frac{u_\ell(r)}{u_\ell(a)}
= 1$ from the right side of Eq. (18) taking the limit term by term, one obtains
a null result, because of Eq. (16).
\vspace{8pt}

In the case of the boundary conditions (5) we can obtain from Eq. (10), that
$$ \eta(r) = \frac{1}{a} (a \frac{d}{dr} - B) \sum_{\lambda = 1}^\infty~
\frac{u_\lambda^\ell(r)u_\lambda^\ell(a)}{k_\lambda^2 - k^2}  ,
\eqno{(19)}
$$

\noindent
where
$$ \eta(r) = [a \frac{du_\ell(r)}{dr} - Bu_\ell(r)]/[a \frac{du_\ell}{dr} - B u_\ell]_{r=a}.
\eqno{(20)}
$$

\noindent
Once again we obtain a null result for $\eta(a) = 1$ by differentiating term
by term and taking the limit term by term of the sum in Eq. (19) and using Eq.
(5).  The explanation for these paradoxes is that either (i) the expansion (7)
or (ii) its 
derivative series, obtained by differentiating the individual terms of
the expansion (7), is not uniformly convergent in the neighborhood of the
surface. Or they may be due to both (i) and (ii).  This difficulty associated with
the expansion (7) has been known for many years
[3,18].
\vspace{8pt}

\noindent
{\bf III.  The Hilbert-Schmidt theorem formulation of the R-matrix theory}.
\vspace{5pt}

Let us rewrite Eq. (3) in an integral form
$$u_\ell(r) = \phi_\ell(r) + (k^2 - \kappa^2) \int_0^a~K_\ell (r,r^\prime)u_\ell(r^\prime)dr^\prime ,
\eqno{(21)}
$$

\noindent
with
$$ K_\ell(r, r^\prime) = - \Biggr\{ \begin{array}{ll}
 \tilde{X}_\ell(r)\tilde{Y}_\ell(r^\prime),&~~~r \leq r^\prime , \\
 \tilde{Y}_\ell(r)\tilde{X}_\ell(r^\prime),&~~~r^\prime \leq r , \\
\end{array}
\eqno{(22)}
$$ 

\noindent
where $\tilde{X}_\ell(r)$ and $\tilde{Y}_\ell(r)$ are regular and
irregular solutions, respectively, of the following equation
$$ - \frac{d^2\psi}{dr^2} + [ \frac{2\mu V(r)}{\hbar^2} + \frac{\ell(\ell + 1}{r^2}] \psi = \kappa^2\psi,
\eqno{(23)}
$$

\noindent
and satisfy the following conditions
$$\everymath={\displaystyle}
\begin{array}{rcl}
 &~& \tilde{X}_\ell(0) = 0 , \\
   & &  \\
 &~& \frac{a}{\tilde{Y}_\ell(a)} (\frac{d \tilde{Y}_\ell}{dr})_{r=a} ~ = B, \\
\end{array}
\eqno{(24)}
$$

\noindent
and 
$$
 \frac{d \tilde{Y}_\ell(r)}{dr} \tilde{X}_\ell(r) - \frac{d \tilde{X}_\ell(r)}{dr}
 \tilde{Y}_\ell(r) = -1 .
$$

\noindent
B is the same as one the introduced in Eq. (5).
$\kappa^2$ is an energy
independent constant satisfying a condition
$$ \kappa^2 \not= k_\lambda^2~~~~~~(\lambda = 1,2,....),
\eqno{(25)}
$$

\noindent
and $\phi_\ell(r)$ is related
to $\tilde{X}_\ell(r)$ by
$$ \phi_\ell(r) = \alpha \tilde{X}_\ell(r),
\eqno{(26a)}
$$

\noindent
where $\alpha$ is an energy dependent constant given by 
$$ \alpha = [(a \frac{du_\ell}{dr} - Bu_\ell(r)]_{r=a}/[a \frac{d \tilde{X}_\ell(r)}{dr} - B \tilde{X}_\ell(r)]_{r=a} .
\eqno{(26b)}
$$

\noindent
The integral equation (21), which is not the Lippmann-Schwinger type equation,
was first introduced in [9] for the $\kappa^2 = 0, B = \infty$ case.  
Eq. (21) has a unique solution, since 
$$ \int_0^a~\int_0^a~K_\ell^2(r,r^\prime)drdr^\prime < \infty ,
$$

\noindent
i.e. $K_\ell(r,r^\prime)$ is completely continuous and self-adjoint kernel [39].
Let $\gamma_\lambda (\lambda = 1,2,...)$ be eigenvalues
of the Hermitian continuous kernel $K_\ell(r,r^\prime)$
$$ u_\lambda^\ell(r) = \gamma_\lambda~\int_o^a~K_\ell(r,r^\prime)u_\lambda^\ell(r^\prime)dr^\prime, 
\eqno{(27a)}
$$

\noindent
with

$$ \gamma_\lambda = k_\lambda^2 - \kappa^2 .    
\eqno{(27b)}
$$

\noindent
As it is well known, the eigenvalues $\gamma_\lambda$ are real, and the functions
$u_\ell(r)$ and $u_\lambda^\ell(r)$ are continuous.  Due to the
Hilbert-Schmidt theorem [39], the following expansion
$$ \int_o^a~K_\ell(r,r^\prime) u_\ell(r^\prime)dr^\prime = ~ \sum_{\lambda = 1}^\infty~\tilde{c}_\lambda^\ell u_\lambda^\ell(r)
\eqno{(28)}
$$

\noindent
converges uniformly and absolutely over $0 \leq r \leq a$, and, if $k^2 \not= k_\lambda^2$, the unique solution $u_\ell(r)$ of the integral equation (21) appears in
the following form of a series which is uniformly and absolutely convergent over $0 \leq r \leq a$
(by Schmidt's formula): 
$$ u_\ell(r)[a \frac{du_\ell}{dr} - Bu_\ell]_{r=a}^{-1} = \frac{\tilde{X}_\ell(r)}{[a \frac{d\tilde{X}_\ell(r)}{dr} - B \tilde{X}_\ell(r)]}_{r=a} + \frac{k^2 - \kappa^2}{a} ~
\sum_{\lambda = 1}^\infty~\frac{u_\lambda^\ell(r)u_\lambda^\ell(a)}{(k_\lambda^2-\kappa^2) (k_\lambda^2 - k^2)} .
\eqno{(29)}
$$

\noindent
If we now define
$$ R^{(B)}(k^2) = R^{(B)}(\kappa^2) + \frac{(k^2-\kappa^2)}{a}~\sum_{\lambda = 1}^\infty~
\frac{(u_\lambda^\ell(a))^2}{(k_\lambda^2 - \kappa^2)(k_\lambda^2 - k^2)} ,
\eqno{(30)}  
$$

\noindent
where
$$ R^{(B)} (\kappa^2) = \frac{\tilde{X}_\ell(a)}{[a \frac{d \tilde{X}_\ell(r)}{dr} -
B \tilde{X}_\ell(r)]}_{r=a}  ,
\eqno{(31)}
$$

\noindent
we find that $R^{(B)}(k^2)$ relates the amplitudes $u_\ell$ to its
derivative on the boundary by the relation (13).  Because the series (29) converges
uniformly and absolutely, the following equation
$$ [\sum_{\lambda = 1}^\infty~\frac{u_\lambda^\ell(r)u_\lambda^\ell(a)}{(k_\lambda^2 - \kappa^2)(k_\lambda^2 - k^2)}]_{r=a} ~= \sum_{\lambda = 1}^\infty~\frac{(u_\lambda^\ell(a))^2}{(k_\lambda^2 - \kappa^2)(k_\lambda^2 - k^2)}
\eqno{(32)}
$$

\noindent
is valid [40], and hence the expansion (29) is free of difficulties encountered
in the expansions given by Eqs. (7) and (10). Series (30) can
be also
obtained from the dispersion formula (11) by separating the energy
independent term $R^{(B)} (\kappa^2)$ in the R-function, Eq. (11), with $\kappa^2$ satisfying condition (25).
Our derivation has shown that   
the dispersion expansion (30) converges absolutely, and exhibits the general
energy-dependence of the R-function.  The expansion given by Eq. (29) is
a main result of this paper.  The proof of the absolute convergence of
the series (30) in case $\kappa^2 = 0$ was given by M. Schiffer and
V. Bargmann.  Their proof is reproduced in [41].  
\vspace{8pt}

\noindent
{\bf IV.  Improving the Convergence}
\vspace{5pt}

In general, the R-function has an infinite number of pole terms.  According
to Courant's minimax considerations, if $V(r)$ is bounded, no $k_\lambda^2$
differs from the corresponding value of $k_\lambda^2$, $(k_\lambda^{(0)})^2$,
for noninteracting
case $(V(r) = 0)$ by more than the bound [3].  Consequently, the general term
of the series (11) for fixed $k^2$ behaves as $1/\lambda^2$ since $(k_\lambda^{(0)})^2
\propto \lambda^2$, while the
general term of the series (30) behaves as $1/\lambda^4$ as $\lambda 
\rightarrow \infty$. For the case of the dispersion formula (11), truncation
of the R-function by a finite number (N) of terms gives
$$ R_N^{(B)} = \frac{1}{a}~\sum_{\lambda = 1}^N~\frac{(u_\lambda^\ell(a))^2}{(k_\lambda^2 - k^2)} 
\eqno{(33)}
$$

\noindent
While for the case of the dispersion formula (30), we have  
$$ R_N^{(B)}(k^2) = R_0^{(N)} + \frac{1}{a}~\sum_{\lambda=1}^N~
\frac{(u_\lambda^\ell(a))^2}{(k_\lambda^2 - k^2)} ,
\eqno{(34)}
$$

\noindent
where
$$ R_0^{(N)} = R^{(B)}(\kappa^2) - \frac{1}{a}~\sum_{\lambda=1}^N~\frac{(u_\lambda^\ell(a))^2}{(k_\lambda^2 - \kappa^2)} .
\eqno{(35)}
$$

The general method of improving the convergence is to separate and sum the slowly
converging parts of the series [42].  It is obvious
that there are many possibilities to obtain a rapid convergence.  For
example, the expansion (30) can be represented in the form
$$
\everymath={\displaystyle}
\begin{array}{rcl}
 R^{(B)}(k^2) &=& R^{(B)}(\kappa^2) + \frac{k^2- \kappa^2}{(k_0^2 - \kappa^2)}
(R^{(B)}(k_0^2) - R^{(B)}(\kappa^2))  \\
& & \\
&+& \frac{(k^2- \kappa^2)(k^2-k_0^2)}{a}~
\sum_{\lambda = 1}^\infty~\frac{(u_\lambda^\ell(a))^2}{(k_\lambda^2 - \kappa^2)(k_\lambda^2 - k^2)(k_\lambda^2 - k_0^2)} 
\end{array}
\eqno{(36)}
$$

\noindent
where $k_0^2$ is an energy independent constant $(k_0^2 \not= \kappa^2, k_0^2 \not=
k_\lambda^2, \lambda = 1,2,...)$.  
Expansion (36) converges much faster than (30) (general term behaves as
$1/\lambda^6)$, and truncation of Eq. (36) by a finite number of terms gives
$$R_N^{(B)}(k^2) = \stackrel{\approx}{R}_0^{(N)} + k^2 R_1^{(N)} + \frac{1}{a}~\sum_{\lambda = 1}^N~
\frac{(u_\ell^\lambda(a))^2}{k_\lambda^2 - k^2} ,
\eqno{(37)}
$$

\noindent
where
$$
\everymath={\displaystyle}
\stackrel{\approx}{R}_0^{(N)} = R^{(B)}(\kappa^2) - \frac{\kappa^2}{k_0^2 - \kappa^2}
(R^{(B)}(k_0^2) - R^{(B)}(\kappa^2)) + \frac{\kappa^2 - k_0^2 - k_\lambda^2}{a}~\sum_{\lambda = 1}^N~\frac{(u_\ell^\lambda(a))^2}{(k_\lambda^2 - \kappa^2)(k_\lambda^2 - k_0^2)} , 
\eqno{(38)}
$$

\noindent
and

$$R_1^{(N)} = \frac{1}{k_0^2 - \kappa^2}(R^{(B)}(k_0^2) - R^{(B)}(\kappa^2)) - \frac{1}{a}~\sum_{\lambda = 1}^N~\frac{(u_\ell^\lambda(a))^2}{(k_\lambda^2 - \kappa^2)(k_\lambda^2 - k_0^2)} .
\eqno{(39)}
$$

\noindent
However, for the case of Eq. (37), we have introduced an additional
parameter $R_1^{(N)}$,  
and we do not expect a weak dependence of Eq. (37) on this parameter
$R_1^{(N)}$.
\vspace{8pt}

To obtain a faster convergence, we introduce a trial potential
$\tilde{V}(r)$ and remove the
corresponding R-function $\tilde{R}^{(B)}(k^2)$ obtained with $\tilde{V}(r)$: 
$$ \tilde{R}^{(B)}(k^2) = \tilde{R}^{(B)}(\kappa^2) + \frac{k^2 - \kappa^2}{a}~\sum_{\lambda = 1}^\infty~\frac{(\tilde{u}_\lambda^\ell(a))^2}{(\tilde{k}_\lambda^2 - \kappa^2)(\tilde{k}_\lambda^2 - k^2)} .
\eqno{(40)}
$$

For the case of the dispersion formula (11), this method has been used in many
papers [11, 15, 18].  For the case of Eq. (30), we have 
$$ \everymath={\displaystyle}
\begin{array}{rcl}
R^{(B)}(k^2) &=& R^{(B)}(\kappa^2) + \tilde{R}^{(B)}(k^2) - \tilde{R}^{(B)}(\kappa^2) \\
& & \\
&+& \frac{k^2 - \kappa^2}{a}~\sum_{\lambda = 1}^\infty~(\frac{(u_\lambda^\ell(a))^2}{(k_\lambda^2 - \kappa^2)(k_\lambda^2 - k^2)} - \frac{(\tilde{u}_\lambda^\ell(a))^2}{(\tilde{k}_\lambda^2 - \kappa^2)(\tilde{k}_\lambda^2 - k^2)}) .
\end{array}
\eqno{(41)}
$$  

\noindent
It can be shown (see Appendix) that the general term of Eq. (41) behaves as $1/\lambda^6$ for any bound
$\tilde{V}(r)$, and hence we expect a weak $\tilde{V}(r)$ dependence for
the following approximation
$$ R_N^{(B)}(k^2) = R_0^{(N)} + g_N(k^2) + \frac{1}{a}~\sum_{\lambda = 1}^N~
\frac{(u_\lambda^\ell(a))^2}{k_\lambda^2 - k^2} ,
\eqno{(42)}
$$

\noindent
where
$$ g_N(k^2) = \tilde{R}^{(B)}(k^2) - \tilde{R}_0^{(N)} - \frac{1}{a}~
\sum_{\lambda = 1}^N~\frac{(\tilde{u}_\lambda^\ell(a))^2}{\tilde{k}_\lambda^2 -
k^2},
\eqno{(43)}
$$
$$ \tilde{R}_0^{(N)} = \tilde{R}^{(B)}(\kappa^2) - \frac{1}{a}~\sum_{\lambda = 1}^N~\frac{(\tilde{u}_\lambda^\ell(a))^2}{\tilde{k}_\lambda^2 - \kappa^2} ,
\eqno{(44)}
$$

\noindent
and $\tilde{u}_\lambda^\ell(r)$ are solutions of Eq. (4) with the trial
potential $\tilde{V}(r)$.  Note that the case of $\tilde{V}(r) = 0$ was
considered in [9, 43].  For practical calculations of the R-matrix with
the approximation
(42) for an incident nucleon, it is possible to use a
simple squared-well potential
$$ \tilde{V}(r) = -V_0 \theta(a - r) ,
\eqno{(45)}
$$

\noindent
where $V_0 = \frac{\hbar^2 K_0^2}{2 \mu}$. The wave number $K_0$ is
independent of the mass number A and is approximately the same for
all nuclei $(K_0 \approx 1 fm)$ [44].
\vspace{12pt}

\noindent
{\bf V.  Summary and Conclusions}
\vspace{5pt}

Using the Hilbert-Schmidt theorem and the integral equation, Eq. (21), we have
reformulated the R-function theory in terms of the expansion given by Eq. (29)
which is uniformly and absolutely convergent
for all values of $0 \leq r \leq a$.  This expansion,
Eq. (29), can be differentiated term by term in the neighborhood of the surface.
Our reformulation solves the existing formal problem of how to formulate the
R-matrix theory without the use of expansions which are not uniformly
convergent.  A possible method for improving the convergence of the R-function
series is given when the series is truncated for practical applications.
\vspace{8pt}

\begin{center}
{\bf Acknowledgments}
\end{center}
\vspace{5pt}

One of the authors (A.L.Z.) acknowledges Drs. L.A.P. Balazs, 
M.S. Goldshtein and S. Khlebnikov for helpful discussions.
\pagebreak

\begin{center}
{\bf Appendix}
\end{center}
\vspace{8pt}

In this Appendix, we show that the general term of Eq (41) behaves
as $1/\lambda^6$.  We rewrite Eq. (4) in a form of the Volterra integral
equation
$$ u_\lambda^\ell(r) = \phi_{0 \lambda}^\ell(r) + \int_0^r~\tilde{K}(r,r^\prime)
u_\lambda^\ell(r^\prime)dr^\prime ,
\eqno{(A.1)}
$$

\noindent
where
$$ \tilde{K}(r,r^\prime) = [\chi_{0 \lambda}^\ell(r)\phi_{0 \lambda}^\ell(r^\prime) -
\phi_{0 \lambda}^\ell(r)\chi_{0 \lambda}^\ell(r^\prime)]\omega(r^\prime) ,
\eqno{(A.2)}
$$

\noindent
with

$$\omega(r) = \frac{2 \mu}{\hbar^2} V(r)
 - [k_\lambda^2 - (k_\lambda^{(0)})^2] .
$$

$\phi_{0 \lambda}^\ell$ in Eq. (A.1) is a regular solution of the Schr\"odinger
equation for noninteracting case
$$ - \frac{d^2 \phi_{0 \lambda}^\ell}{dr^2} + \frac{\ell(\ell + 1)}{r^2}
\phi_{0 \lambda}^\ell(r) = (k_\lambda^{(0)})^2 \phi_{0 \lambda}^\ell (r) ,
\eqno{(A.3)}
$$

\noindent
satisfying the R-matrix boundary conditions (5) and the orthonormality
conditions (6).  $\chi_{0 \lambda}^\ell$ in Eq. (A.2) is the irregular
solution of Eq. (A.3)

$$ \chi_{0 \lambda}^\ell(r) = \phi_{0 \lambda}^\ell(r)
\int_0^r [\phi_{0 \lambda}^\ell(x)]^{-2}dx ,
$$

\noindent
and $k_\lambda^2$ is defined from the condition
$$ \int_0^a~\chi_{0 \lambda}^\ell(r) \omega(r)u_\lambda^\ell(r)dr = 0 .
\eqno{(A.4)}
$$
\vspace{8pt}

For any bound and continuous $V(r), \tilde{K}(r, r^\prime)$ is also
continuous and bound, and hence the Neumann series (iteration series)
$$ u_\lambda^\ell(r) = \sum_{p=0}^\infty~(\tilde{K}^p \phi_{0 \lambda}^\ell)(r)
\eqno{(A.5)}
$$

\noindent
converges uniformly and absolutely over $0 \leq r \leq a$ [39], where
$\tilde{K}^p$ is a product of the operators
$\tilde{K}$ and the function $\tilde{K}(r,r^\prime)$  is the kernel
of the linear integral operator $\tilde{K}$.
\vspace{5pt}

>From Eq. (A.5), we can obtain
$$
 \lim_{\lambda \rightarrow \infty}~u_\lambda^\ell(a)
= \phi_{0 \lambda}^\ell(a) + O(\frac{1}{\lambda^2}) ,
\eqno{(A.6)}
$$

\noindent
and hence the general term of Eq. (41) behaves as $1/\lambda^6$ for any
bound and continuous $V(r)$ and $\tilde{V}(r)$.
\vspace{8pt}

We note an important fact that we do not need ``smallness" of V(r) for the
convergence of the Neumann series, Eq. (A.5),
in contrast to the conventional perturbation expansion.
\pagebreak

\begin{center}
{\bf References}
\end{center}
\vspace{8pt}

\begin{enumerate}

\item E. P. Wigner and L. Eisenbud, Phys. Rev. {\bf 72}, 29 (1947).

\item G. Breit and W. G. Bouricius, Phys. Rev.{\bf 75}, 1029 (1949).

\item A. M. Lane and R. G. Thomas, Rev. Mod. Phys. {\bf 30}, 257 (1958).

\item C. Bloch, Nucl. Phys. {\bf 4}, 503 (1957).

\item H. Feshbach and E. L. Lomon, Phys. Rev. {\bf 102}, 891 (1956);
Ann. Phys. (N.Y.) {\bf 29}, 19 (1964).

\item Y. E. Kim and A. Tubis, Phys. Rev. {\bf C1}, 414 (1970); Phys.
Rev. {\bf C2}, 2118 (1970); Phys. Rev. Lett. {\bf 31}, 952 (1973).

\item V. N. Efimov and H. Schulz, Sov. J. Part. Nucl. {\bf 7}, 349 (1976).

\item R. L. Jaffe and F. E. Low, Phys. Rev. {\bf D19}, 2105 (1979).

\item A. Abdurakhmanov, A. L. Zubarev, A. Sh. Latipov, and M. Nasyrov,
Sov. J. Nucl. Phys. {\bf 46}, 217 (1987).

\item V. A. Babenko, N. M. Petrov and A. G. Sitenko, Can. J. Phys.
{\bf 70}, 252 (1991).

\item P.J.A. Buttle, Phys. Rev. {\bf 160}, 719 (1967).

\item F. C. Barker, H. J. Hay and P. B. Treacy, Aust. J. Phys. {\bf 21},
239 (1968).

\item F. C. Barker, Aust. J. Phys {\bf 25}, 341 (1972); Nucl. Phys.
{\bf A588}, 693 (1995).

\item F. C. Barker and T. Kajino, Aust. J. Phys. {\bf 44}, 369 (1991).

\item S. E. Koonin, T. A. Tombrello, and G. Fox, Nucl. Phys. {\bf A220},
221 (1974).

\item H. D. Knox, D. A. Resler and R. O. Lane, Nucl. Phys. {\bf A466},
245 (1987).

\item G. M. Hale, R. E. Brown, and N. Jarmie, Phys. Rev. Lett. {\bf 59},
763 (1987).

\item P. G. Burke and W. D. Robb, Adv. in Atomic and Molecular Phys. {\bf 11}, 143 (1975).

\item K. A. Berrington, P. G. Burke, M. Le Dournef, W. D. Robb, K. T.
Taylor, and Lan Vo Ky, Comput. Phys. Commun. {\bf 14}, 346 (1978).

\item K. A. Berrington, P. G. Burke, K. Butler, M. J. Seaton, P. Y. Storey,
K. T. Taylor, and Yan Yu, J. Phys. {\bf B20}, 6379 (1987).

\item A.Lisini, P. G. Burke, and A. Hilbert, J. Phys. {\bf B23}, 3767
(1990).

\item Lan Vo Ky, N. E. Saraph, W. Eissner, Z. W. Liu, and H. P. Kelly, 
Phys. Rev. {\bf A46}, 3945 (1992).

\item B. Schneider, Chem. Phys. Lett. {\bf 31}, 237 (1975).

\item B. Schneider, Phys. Rev. {\bf A11}, 1957 (1975).

\item P. G. Burke, I. Mackey, and I. Shimamura, J. Phys. {\bf B10},
2497 (1977).

\item B. I. Schneider, M. LeDourneuf, and P. G. Burke, J. Phys.
{\bf B12}, L365 (1979).

\item B. I. Schneider, M. LeDourneuf, and Vo Ky Lan, Phys. Rev.
Lett. {\bf 43}, 1926 (1979).

\item B. I. Schneider, {\it Electron-Atom and Electron-Molecule
Collisions}, Ed. by J. Hinze (Plenum Publishing Corporation, 1983),
p. 121.

\item P. G. Burke and C. J. Noble, Comments At. Mol. Phys. {\bf 18},
181 (1986).

\item A. M. Lane and D. Robson, Phys. Rev. {\bf 178}, 1715 (1968).

\item R. S. Oberoi and R. K. Nesbet, Phys. Rev. {\bf A8}, 2115 (1973).

\item R. S. Oberoi and R. K. Nesbet, Phys. Rev. {\bf A9}, 2804 (1974).

\item L. Schlessinger and G. L. Payne, Phys. Rev. {\bf A10}, 1559 (1974).

\item U. Fano and C. M. Lee, Phys. Rev. Lett. {\bf 31}, 1573 (1973).

\item C. M. Lee, Phys. Rev. {\bf A10}, 584 (1974).

\item R. K. Nesbet, {\it Variational Method in Electron-Atom Scattering
Theory} (Plenum Press, New York, 1980).

\item A. G. Sitenko, {\it Theory of Nuclear Reactions} (World Scientific, Singapore,
1990).

\item L. M. Kiernan, M. K. Lee. B. F. Sonntag, P. Zimmerman, J. T. Costello,
E. T. Kennedy, A. Gray, and Lan Vo Ky, J. Phys. {\bf B29}, L21 (1996).

\item V. S. Vladimirov, {\it Equation of Mathematical Physics} (Marcel Dekker,
Inc., New York, 1971);  A.N. Kolmogorov and S.V. Fomin, {\it Elements of the
Theory of Functions and Functional Analysis} (Graylock Press, Albany, N.Y.,
1961).

\item K. Knopp, {\it Theory and Application of Infinite Series} (Hefner Publishing
Company, New York 1947).

\item E. Wigner, Ann. Math. {\bf 53}, 36 (1951).

\item L.V. Kantorovich and V.I. Krylov, {\it Approximate Methods of Higher
Analysis} (Interscience Publisher, Inc., N.Y., 1958).

\item V.A. Babenko and N. M. Petrov, Sov. J. Nucl. Phys. {\bf 45}, 1004 (1987).

\item J.. M. Blatt and V.F. Weisskopf, {\it Theoretical Nuclear Physics} (John
Wiley and Sons, N.Y., 1963).

\end{enumerate}

\end{document}